\documentclass[prc,preprint,showpacs,superscriptaddress,floatfix,amsmath,amssymb,nofootinbib]{revtex4}
\usepackage{graphicx}
\usepackage{color}
\usepackage{mathrsfs}

\newcommand{\beq}{\begin{equation}}
\newcommand{\eeq}{\end{equation}}
\newcommand{\beqn}{\begin{eqnarray}}
\newcommand{\eeqn}{\end{eqnarray}}

\newcommand{\ls}{\left[}
\newcommand{\rs}{\right]}

\begin{document}%
\title{Signature splitting in $^{173}$W with triaxial particle rotor model
}

\author{B. Qi }
 \affiliation{School of Physics, and State-Key Lab. Nucl. Phys.
\& Tech., Peking University, Beijing 100871, China}

\author{S.Q. Zhang }\thanks{sqzhang@pku.edu.cn}
 \affiliation{School of Physics, and State-Key Lab. Nucl. Phys. \& Tech., Peking University, Beijing 100871, China}
 \affiliation{Institute of Theoretical Physics, Chinese Academy of Sciences, Beijing,100080, China}

\author{S.Y. Wang }
 \affiliation{ Department of Space Science and Applied Physics, Shandong University at Weihai, Weihai 264209, China}

\author{J. Meng } \thanks{mengj@pku.edu.cn}

 \affiliation{School of Physics, and State-Key Lab. Nucl. Phys. \& Tech., Peking University, Beijing 100871, China}
 \affiliation{Institute of Theoretical Physics, Chinese Academy of Sciences, Beijing,100080, China}
 \affiliation{Department of Physics, University of Stellenbosch, Stellenbosch, South Africa}
 \affiliation{Center of Theoretical Nuclear Physics, National Laboratory of Heavy Ion Accelerator, Lanzhou 730000, China}


\begin{abstract}
A particle rotor model with a quasi-neutron coupled with a
triaxially deformed rotor is applied to study signature splitting
for bands with intruder orbital $\nu7/2^{+} [633]$ and non-intruder
orbital $\nu5/2^{-}[512]$ in $^{173}$W. Excellent agreement with the
observed energy spectra has been achieved for both bands. Signature
splitting for band $\nu7/2^{+} [633]$, and band $\nu5/2^{-}[512]$
before the onset of signature inversion, is satisfactorily
reproduced by introducing the $\gamma$ degree of freedom. The phase
and amplitude of signature splitting in band $\nu5/2^{-}[512]$ is
attributed to strong competition between $2f_{7/2}$ and $1h_{9/2}$
components. However, the explanation of signature inversion in band
$\nu5/2^{-}[512]$ self-consistently is beyond the present one
quasi-neutron coupled with a triaxially deformed rotor.
\end{abstract}


\maketitle

\section{Introduction}

Signature splitting in rotational bands, and especially signature
inversion, has attracted much attention for decades. Signature
$\alpha$ is a quantum number describing the symmetry associated with
the rotation of $180^{\circ}$ about one of the three principal axes.
The spin sequences for a band with signature $\alpha$ are $ I =
\alpha+ 2n (n= 0,1,...)$, with $\alpha$ = 0 or 1 in even-mass nuclei
and $\alpha=\pm \dfrac{1}{2}$ in odd-mass nuclei. In general,
signature partner bands are not energetically equivalent. One band
is favored, i.e., lower in energy, whereas the other one is
separated by the so-called signature splitting, which is due to the
Coriolis interaction. Signature splitting is usually characterized
by $ S(I)=\ls E(I)-E(I-1)\rs-\frac{1}{2}\ls E(I+1)-E(I)+
E(I-1)-E(I-2)\rs$, which appears as a typical staggering curve. An
interesting phenomenon is that in several cases the energetically
favored and unfavored signature bands may cross each other and
interchange their role, i.e., the signature splitting changes its
staggering phase with spin $I$. This phenomenon is known as
signature inversion~\cite{Bengtsson84}. For the configuration of
one-quasiparticle (1-qp) in high-$j$ orbitals, the favored signature
is obtained by a simple rule $\alpha_f=(-1)^{j-1/2}
\dfrac{1}{2}$~\cite{Stephens75}. However, for a 2-qp configuration
in odd-odd nuclei, the expectative favor band with
$\alpha_f=(-1)^{j_p-1/2} \dfrac{1}{2}+ (-1)^{j_n-1/2} \dfrac{1}{2}$
may become energetically unfavored. Similar effects have been also
observed for some 3-qp bands in odd-$A$ nuclei. One may also refer
to such phenomenon as signature inversion.

Signature inversions have been observed systematically in regions of
mass number $A\sim$80, $A\sim$130 and $A\sim$160. See for example
Refs.~\cite{Liu95,Liu96,Plettner00, Garcia01, Singh07}. For 2-qp
bands of odd-odd nuclei, signature inversions generally take place
in low-spin regions including the bandhead states. For odd-$A$
nuclei, they occur at higher spins than the first backbendings,
i.e., in 3-qp bands~\cite{Tajima94}. Possible mechanisms for
signature inversion have been proposed, such as
triaxiality~\cite{Bengtsson84,Ikeda89, Hamamoto83,Plettner00}, band
crossing~\cite{Plettner00, Hara92}, the n-p
interaction~\cite{Hamamoto862, Semmes91, Cederwall92, Tajima94,
Zheng01}, QQ pairing~\cite{Sautula96, XuFR00}, and the drift of the
rotational axis in triaxial nuclei~\cite{Gao06}.

In $A\sim$160 mass region, signature inversions have been
extensively investigated for the bands based on the configurations
$\pi h_{11/2}\otimes\nu i_{13/2}$ and $\pi h_{9/2}\otimes\nu
i_{13/2}$. Recently, signature inversions in the odd-odd nucleus
$^{174}$Re were found not only for the two configurations mentioned
above~\cite{ZhangYHCPL05} but also for a new band based on the
configuration $\pi h_{9/2}\otimes\nu  5/2^- [512]$
~\cite{ZhangYHCPL07}. The authors of Ref.~\cite{ZhangYHCPL07}
suggested that the $\nu 2f_{7/2}$ and $\nu 1h_{9/2}$ configuration
admixture may be a possible reason for signature inversion in the
band $\pi h_{9/2}\otimes\nu 5/2^- [512]$. The investigation of the
neighboring odd-$N$ nuclei $^{173}$W and $^{175}$Os would be
beneficial to the interpretation thereof. A recent experiment on
$^{173}$W was performed~\cite{ZhangYH06} and three former observed
rotational bands~\cite{Walker78}, whose configurations are
respectively proposed as $\nu 5/2^{-} [512]$, $\nu 7/2^{+} [633]$
and $\nu 1/2^{-} [521]$,  were extended to higher spin states. It is
interesting to note that signature inversion for band $\nu 5/2^{-}
[512]$ is observed at spin $\frac{35}{2}\hbar$. Signature inversion
in band $\nu5/2^-[512]$  is also observed in other isotones of
$^{173}$W, i.e., $^{171}$Hf~\cite{Cullen00} and
$^{175}$Os~\cite{Fabricius90}.

The aim of the present work is to investigate signature splitting in
$^{173}$W in the triaxial particle rotor model. The band $\nu
7/2^+[633]$ in $^{173}$W and similar bands in $^{175}$W have been
investigated by assuming a quasi-particle in a deformed Nilsson
potential coupled with an axial symmetric rotor~\cite{Walker78}. In
this paper, the effect of triaxiality will be investigated. The
model will be briefly introduced in Sec. II. Signature splitting of
bands $\nu 5/2^-[512]$  and $\nu7/2^+[633]$ will be discussed in
detail in Sec. III and a brief summary will be given in Sec. IV.

\section{Formalism}

The particle rotor model (PRM) adopted here is same as in
Refs.~\cite{Meng96,Meng97} and has been extensively used in the
investigation of the chirality in atomic nuclei~\cite{Peng03,
WangSY07, Zhang07, WangSY08}. The model Hamiltonian can be expressed
as,
 \beq
 \label{eq:hamiltonian}
    H=H_\textrm{coll}+ H_\textrm{intr}.
 \eeq
The collective Hamiltonian with a triaxial rotor can be written as
 \beq
 H_\textrm{coll}
 = \sum_{i=1}^{3} \frac{\hat{R}_{i}^2}{2{\cal J}_i}
 = \sum_{i=1}^{3}
\frac{(\hat{I}_{i}-\hat{j}_{i})}{2{\cal J}_i} ,
 \label{eq:hcoll}
 \eeq
where $\hat{R}_i, \hat{I}_{i}, \hat{j}_{i}$ respectively denote the
angular momentum operators for the core, nucleus, as well as the
valence nucleon.  The moments of inertia for irrotational flow are
given by
 \beq {\cal J}_i
    = \frac{4}{3}{\cal J}_0 \sin^2(\gamma - \frac{2\pi}{3}i), ~~~(i =
    1,2,3),
 \eeq
where ${\cal J}_0$ depends on the quadrupole deformation
$\varepsilon_2$ and the nuclear mass $A$~\cite{Meyer75}, while
$\gamma$ denotes the degree of triaxiality.

The intrinsic Hamiltonian for valence nucleon is
 \beq
 H_\textrm{intr}
 = H_\textrm{sp}+H_\textrm{pair}
 = \sum_{\nu >0 }(\varepsilon_\nu-\lambda)
   (a_\nu^{+}a_\nu + a_{\overline{\nu}}^{+}a_{\overline{\nu}} )
 -\frac{\Delta}{2}\sum_{\nu >0}
 (a_{\nu}^{+}a_{\overline\nu}^{+}+a_{\overline{\nu}}{a_\nu}),
 \eeq
where $\lambda$ denotes the Fermi energy, $\Delta$ the pairing gap
parameter, and $|\overline{\nu}\rangle$ the time-reversal state of
$|{\nu}\rangle$. The single particle states $|{\nu}\rangle$ and
corresponding energies $\varepsilon_\nu$ are obtained by
diagonalizing the Hamiltonian $H_\textrm{sp}$. Similar to
Refs.~\cite{Larsson78,Ragnarsson88}, we employ the Nilsson type
Hamiltonian,
 \beqn\label{eq:hsp}
  H_\textrm{sp}
  &=&(\frac{p^{2}}{2m}
     +\frac{1}{2}m\omega_{0}^{2}\rho^2)
     -\frac{2}{3}\varepsilon_{2}\sqrt{\frac{4\pi}{5}}\hbar\omega_0 \rho^2
     \ls \cos \gamma Y_{20}+ \frac{1}{\sqrt{2}} \sin \gamma ( Y_{22}+Y_{2-2})\rs\nonumber\\
  &&-\kappa \hbar \stackrel{\circ}{\omega _{0}} \{2 \vec{l}\cdot \vec{s}
     +\mu(\vec{l}^2- \langle\vec{l} ^2\rangle_{N} ) \}.
 \eeqn

The single particle states are thus written as
 \beq
 {a}^{+}_{\nu}| 0 \rangle
   =\sum_{Nl j\Omega}c_{Nl j\Omega}^{(\nu)}\psi^{Nl}_{j\Omega},~~~~~
 {a}^{+}_{\overline{\nu}}| 0 \rangle
   = \sum_{Nl j\Omega}(-1)^{j-\Omega}c_{Nl j\Omega}^{(\nu)}\psi^{Nl}_{j-\Omega},~
  \eeq
where $\Omega$ is the projection of the single-particle angular
momentum ${\hat j}$ along the 3-axis and can be restricted to the
values $\cdots, -7/2, -3/2, +1/2, +5/2, \cdots$ due to time-reversal
degeneracy~\cite{Larsson78,Ragnarsson88}.

To obtain the PRM solutions, the total
Hamiltonian~(\ref{eq:hamiltonian}) must be diagonalized in a
complete basis space, which couples the rotation of the inert core
with the intrinsic wave functions of the valence nucleons. When
pairing correlations are neglected, one can construct the so-called
strong coupling basis as
  \beqn\label{eq:base}
  |IMK\nu\rangle &=&\sqrt{\frac{1}{2}}
  \sqrt{\frac{2I+1}{8\pi^2}}\ls
  D_{M,K}^{I} a^{+}_{\nu}
  |0\rangle
   +(-1)^{I-K}D_{M,-K}^{I} a^{+}_{\overline{\nu}}
  |0\rangle
  \rs\nonumber\\
 {\rm for~} K&=&\cdots, -7/2, -3/2, +1/2, +5/2, \cdots.
 \eeqn
 The restriction on values of $K$ is due to the fact that the basis
states are symmetrized under the point group $D_2$, which leads to
$K-\Omega$ being an even integer~\cite{Larsson78}. The matrix
elements of the Hamiltonians given by Eqs.~(\ref{eq:hcoll}) and
(\ref{eq:hsp}), can be evaluated in the basis (\ref{eq:base}), and
thereafter diagonalization yields eigenenergies and eigenstates of
the PRM Hamiltonian.

To include pairing effects in the PRM, one should replace the single
particle state $a^{+}_{\nu}| 0\rangle$ in the basis
states~(\ref{eq:base}) with the BCS quasiparticle state
$\alpha^{+}_{\nu}|\tilde{0}\rangle$ to obtain a new expansion basis,
where $|\tilde{0}\rangle$ is the BCS vacuum state. The quasiparticle
operators $\alpha_{\nu}^{+}$ are given by
 \beq
 \left(
\begin{array}{c}
  \alpha_{\nu}^{+} \\
  \alpha_{\overline{\nu}} \\
\end{array}
\right) =\left(\begin{array}{cc}
  u_\nu & -v_\nu \\
  v_\nu & u_\nu \\
\end{array} \right)~~
\left(
\begin{array}{c}
  a_{\nu}^{+} \\
  a_{\overline{\nu}} \\
\end{array}\right),
 \eeq
where $u_\nu^2 + v_\nu^2 = 1$. In this new basis, the wave functions
of the PRM Hamiltonian are written as
 \beq
 |IM \rangle = \sum_{K,\nu} C_{\nu}^{IK} |IMK\nu\rangle ,
 \eeq
in which $\nu$ represents the quasiparticle states
$\alpha^{+}_{\nu}|\tilde{0}\rangle$ instead of
$\alpha^{+}_{\nu}|0\rangle$. Furthermore, single-particle energies
$\varepsilon_{\nu }$ should be replaced by quasiparticle energies
 $
 \varepsilon'_{\nu}= \sqrt{(\varepsilon_{\nu} -
 \lambda)^2+ \Delta^2}.
 $
 The total Hamiltonian then becomes:
 \beq H = H_{\rm coll} +
 \sum_{\nu}\varepsilon'_{\nu}(\alpha_{\nu}^{+}\alpha_{\nu}
    + \alpha_{\overline{\nu}}^{+}\alpha_{\overline{\nu}}).
 \eeq
To construct the matrix of the above Hamiltonian, in comparison with
the case excluding pairing, each single-particle matrix element
needs to be multiplied by a pairing factor
$u_{\mu}u_{\nu}+v_{\mu}v_{\nu}$~\cite{Meyer75,Ragnarsson88}. The
occupation factor $v_\nu$ of the state $\nu$ is given by
 \beqn
  v^2_\nu
  =\frac{1}{2}
   \left[1-\frac{\varepsilon_\nu-\lambda}{\varepsilon'_{\nu}}\right].
 \eeqn

\section{Results and Discussion}

In the present calculations, the values of $\kappa$ and $\mu$ in the
Nilsson type Hamiltonian~(\ref{eq:hsp}) are taken from
Ref.~\cite{Bengtsson85}, i.e., $\kappa=0.062$ and $\mu=0.43$ for the
main oscillator quantum number $N=5$, $\kappa=0.062$ and $\mu=0.34$
for $N=6$. The quadrupole deformation parameter $\varepsilon_2$
takes a value of 0.24 according to Ref.~\cite{Moller95}. The
triaxiality parameter $\gamma$ is to be adjusted by the signature
splitting. An off-diagonal Coriolis attenuation parameter $\xi$ is
introduced to reproduce the experimental energy
spectra~\cite{Ragnarsson88}, and a variable moment of inertia is
used when necessary, $~{\cal J}_0(I)={\cal J}_0
\sqrt{1+bI(I+1)}$~\cite{Wu87}. The neutron Fermi energy $\lambda_n$
is taken to be the energy of the single-particle level occupied by
the last neutron. The paring gap parameter $\Delta$ is given by the
empirical value of 0.81 MeV.

For the axially deformed case, the orbitals are usually denoted by
the Nilsson quantum number $\Omega^{\pi}[Nn_3\Lambda]$. For the
triaxially deformed case, as $\Omega$ is not a good quantum number,
the number $\nu$ is used to denote the single-particle state
according to the sequence of the energy. For convenience, the
Nilsson quantum number is also used to denote approximately the
single particle state in the triaxially deformed case. From the
standard Nilsson single particle level diagram,  the last neutron of
$^{173}$W lies in the orbital $1/2^-[521]$ at $\varepsilon_2=0.24$,
with the orbitals $7/2^{+}[633]$, $5/2^-[512]$ nearby, which is
consistent with the configurations proposed for the three observed
bands~\cite{Walker78}. Furthermore, the calculated energy difference
between the states $7/2^{+}[633]$ and $5/2^-[512]$ is 67.6 keV at
$\varepsilon_2=0.24$, which agrees with the observed bandhead energy
difference 85.5 keV in $^{173}$W~\cite{ZhangYH06}.

In Table~\ref{Table:1}, seven positive-parity orbitals and seven
negative-parity orbitals near the Fermi level adopted for band $\nu
7/2^+[633]$ at $\gamma=15^\circ$, and band $\nu 5/2^-[512]$ at
$\gamma=12^\circ$ in the present triaxial PRM calculations are
listed.  The neutron Fermi energy $\lambda_n$ is $51.46$ MeV for
$\gamma=15^\circ$, and $51.51$ MeV for $\gamma=12^\circ$. The above
mentioned orbitals are used to couple with the core in the
calculations for the positive-parity and negative-parity bands of
$^{173}$W respectively. Their approximate Nilsson quantum numbers,
single particle energies, and main components expanded in the basis
$|Nlj\Omega\rangle$ are also listed. One can see that the main
components for the positive parity orbitals near the Fermi level
belong to the $i_{13/2}$ sub-shell components, while for the
negative-parity band there is a strong mixture of $1h_{9/2}$ and
$2f_{7/2}$ sub-shells.

In the following, using the triaxial particle rotor model, we will
focus our discussion on signature splitting in the bands
$\nu7/2^{+}[633]$ and $\nu5/2^-[512]$ .

\subsection{Band $\nu7/2^{+}[633]$}

\begin{center}
\begin{figure*}[h!]
\centering
\includegraphics[height=10cm]{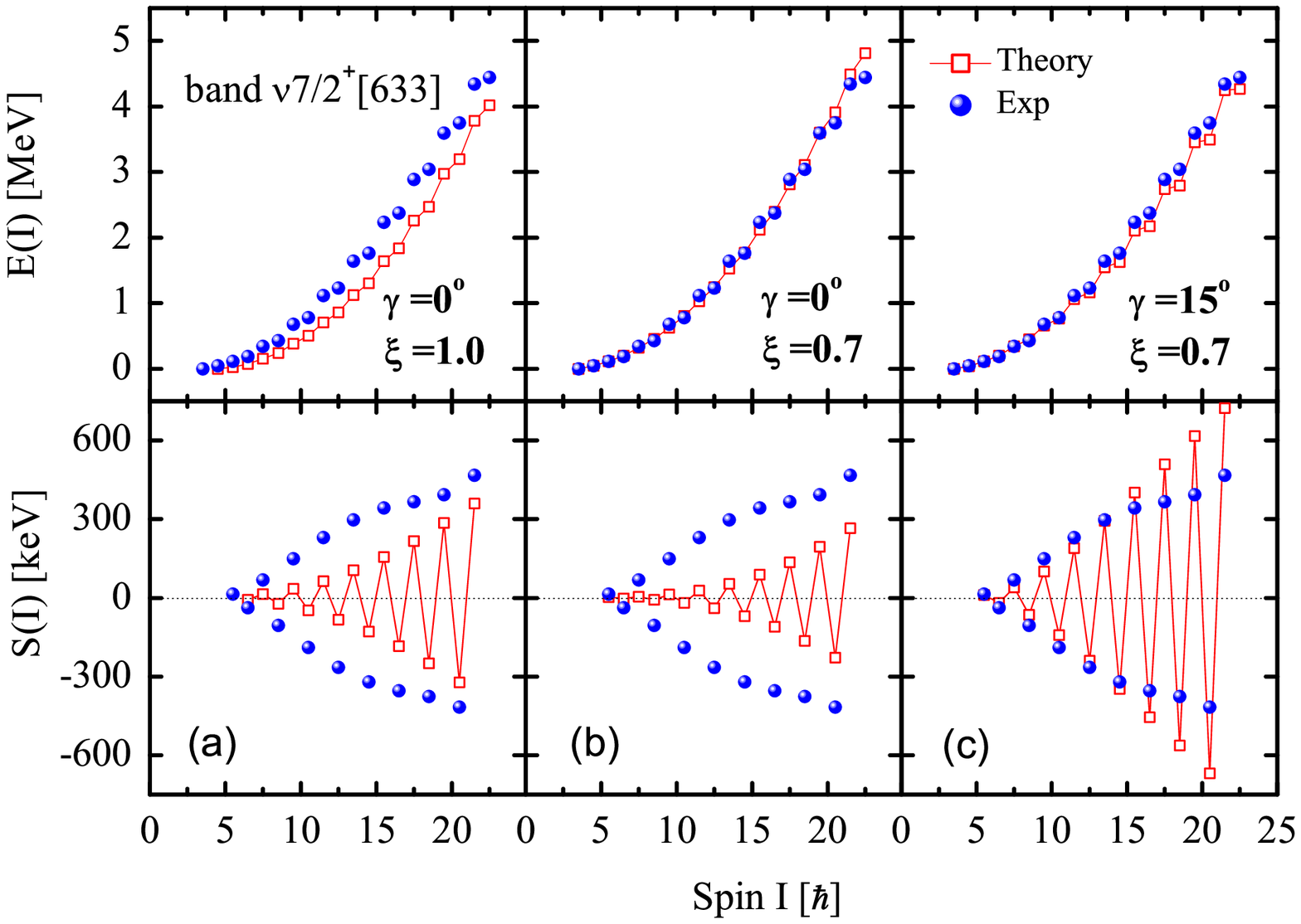}
\caption{(color online) The energy spectra $E(I)$ and signature
splitting $S(I)$ for the band $\nu7/2^{+}[633]$ in $^{173}$W based
on the PRM (red squares) in comparison with the data (blue
points)~\cite{ZhangYH06}. In the calculations, $\varepsilon_2=0.24$,
${\cal J}_0 = 40$ MeV$^{-1}\hbar^2$ and $b=0$. The pairing parameter
$\Delta_{n} = 0.81$ MeV. The Fermi energy $\lambda_n$, triaxial
deformation parameter $\gamma$ and the Coriolis attenuation factor
$\xi$ in the left (a), middle (b) and right(c) panels are
respectively, $\lambda_n = 51.41$ MeV, $\gamma=0^{\circ}$, and
$\xi=1.0$ (a); $\lambda_n = 51.41$ MeV, $\gamma=0^{\circ}$, and
$\xi=0.7$ (b); $\lambda_n = 51.46$ MeV, $\gamma=15^{\circ}$ and
$\xi=0.7$ (c).} \label{fig:fig1}
\end{figure*}
\end{center}

The calculated energy spectra $E(I)$, signature splitting $S(I)$ for
the band $\nu7/2^{+}[633]$ and subsequent comparison with data are
illustrated in Fig.~\ref{fig:fig1}. Firstly, the moment of inertia
${\cal J}_0$ is adjusted to reproduce the energy spectra by
switching off the triaxiality and Coriolis attenuation. As shown in
Fig.~\ref{fig:fig1}a, where the best fitting by adjusting ${\cal
J}_0$ alone is given, the calculation gives the lowest state is at
$I = \frac{9}{2}\hbar$ instead of $I = \frac{7}{2}\hbar$.
Furthermore, obvious deviations with the data for both the energy
spectra and the amplitude of signature splitting can be seen.

Secondly, as in Ref.~\cite{Ragnarsson88}, the Coriolis attenuation
$\xi = 0.7$ is introduced as shown in Fig.~\ref{fig:fig1}b. The
calculation with Coriolis attenuation can give correctly the lowest
state at $I = \frac{7}{2}\hbar$ and achieve better agreement with
the energy in almost whole spin region. However, the amplitude of
signature splitting is too small to reproduce the data.

With the triaxiality deformation switched on and the corresponding
Fermi energy altered as the energy of the level occupied by the last
neutron, good agreement with the amplitude of signature splitting is
achieved, as shown in Fig.~\ref{fig:fig1}c. It is interesting to
note that the triaxiality improves the amplitude of the signature
splitting while does not ruin the agreement with the energy in the
whole spin region.

From Figs.~\ref{fig:fig1}a - \ref{fig:fig1}c, the important roles of
the Coriolis attenuation and $\gamma$ degree of freedom are shown
explicitly to reproduce the energy spectra $E(I)$ and signature
splitting $S(I)$ for the band $\nu7/2^{+}[633]$. A further check was
done and it is confirmed that by adjusting the $\gamma$ alone
without the Coriolis attenuation, the energy spectra $E(I)$ for the
whole spin region cannot be reproduced.

\begin{center}
\begin{figure*}[h!]
\centering
\includegraphics[width=12cm]{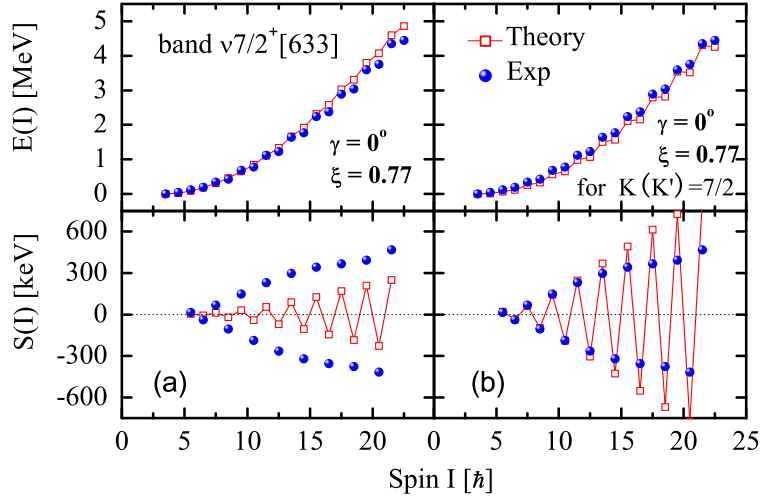}
\caption{(color online) The energy spectra $E(I)$ and signature
splitting $S(I)$ for the band $\nu7/2^+[633]$ in $^{173}$W based on
the PRM (red squares) without triaxiality in comparison with the
data (blue points)~\cite{ZhangYH06}. In the calculations, the same
as Fig. 9 in Ref.~\cite{Walker78}, we have chosen
$\varepsilon_2=0.23, \gamma=0^{\circ}$, and $\hbar^2/2{\cal J} =
A_1+A_2[I(I+1)-K^{2}]$ with $A_1=17.0$ keV and $A_2= -6.7$ eV. In
the left panels (a), the attenuation factor $\xi =0.77$ is
introduced for the off-diagonal Coriolis matrix elements for all $K,
K'$; while in the right panels (b), the attenuation factor $\xi
=0.77$ is introduced only for the matrix elements associated with
the lowest single-quasiparticle state (i.e., $K$ or $K'=7/2$), the
same as in Ref.~\cite{Walker78}.} \label{fig:fig2}
\end{figure*}
\end{center}

In Ref.~\cite{Walker78}, the signature splitting for the band
$\nu7/2^{+}[633]$ has been reproduced without involving triaxiality.
To clarify this point£¬the calculations with the same parameters as
in Ref.~\cite{Walker78} are performed and the results are shown in
Fig.~\ref{fig:fig2}. If the Coriolis attenuation is taken into
account for the off-diagonal Coriolis matrix elements for all $K$,
$K'$, good agreement with the energy can be achieved but not the
amplitude of signature splitting, as shown in Fig.~\ref{fig:fig2}a.
In order to reproduce both the energy and the amplitude of signature
splitting without triaxiality, in Ref.~\cite{Walker78}, the Coriolis
attenuation has to be taken only for the matrix elements associated
with the single-quasiparticle state (i.e., $K$ or $K'=7/2$), as
shown in Fig.~\ref{fig:fig2}b.

\begin{center}
\begin{figure*}[h!]
\centering
\includegraphics[width=8cm]{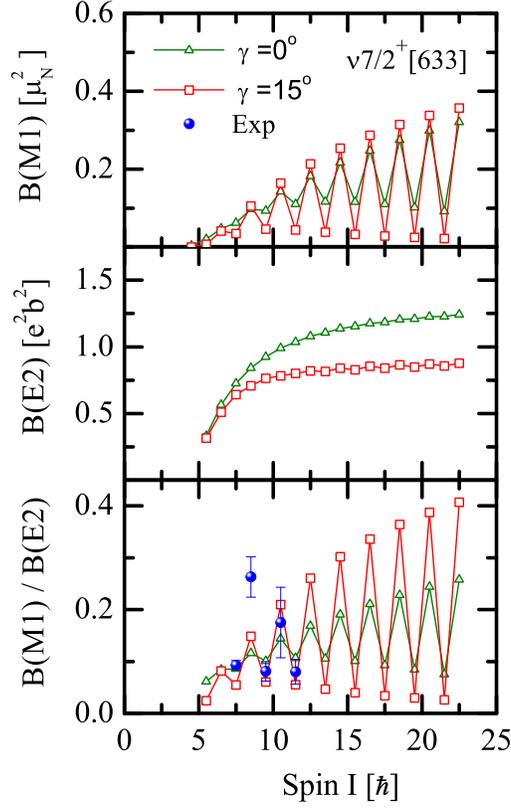}
\caption{(color online) $B(M1)$, $B(E2)$ and $B(M1)/B(E2)$ values
for the band $\nu7/2^{+}[633]$ in $^{173}$W based on the PRM with
$\gamma=0^\circ$ (olive triangles) and $\gamma=15^\circ$ (red
squares) in comparison with the available data (blue
points)~\cite{Walker78}. Other parameters are the same as
Fig.~\ref{fig:fig1}b and Fig.~\ref{fig:fig1}c. In the calculation of
electromagnetic transitions, the empirical intrinsic quadrupole
moment $Q_0= $6 eb, the collective gyromagnetic factor $g_R=0.43$,
and the gyromagnetic factor of quasineutron $g_n=g_l\pm
(g_s-g_l)/(2l+1)$ with $g_l=0, g_s=-2.68$ have been adopted.}
\label{fig:fig3}
\end{figure*}
\end{center}

To show the influence of the triaxiality parameters on the
electromagnetic properties in the band $\nu7/2^{+}[633]$, the
calculated $B(M1)$, $B(E2)$ and $B(M1)/B(E2)$ values are illustrated
and compared with the data  available in Fig.~\ref{fig:fig3}. The
formulas for $B(M1)$ and $B(E2)$ are the same as in
Refs.~\cite{Zhang07,Meyer75}. In the calculations, the empirical
intrinsic quadrupole moment $Q_0=$6 eb, the collective gyromagnetic
factor $g_R=0.43$, and the gyromagnetic factor of quasineutron
$g_n=g_l\pm (g_s-g_l)/(2l+1)$ with $g_l=0, g_s=-2.68$ have been
adopted. The calculated $B(M1)$ values exhibit a staggering. Just as
$S(I)$, the amplitudes of the $B(M1)$ staggering are larger for
$\gamma=15^\circ$ than that for $\gamma=0^\circ$. While the $B(E2)$
values for triaxial deformed case are suppressed in comparison with
the axial deformed case. Therefore, the calculated $B(M1)/B(E2)$
staggering for this band can be enhanced obviously by introducing
triaxiality. The calculated $B(M1)/B(E2)$ staggerings with
triaxiality achieve better agreement with the data
available~\cite{Walker78} than the ones without triaxiality.

We also investigate the main components of the wave functions for
spins $I = \frac{7}{2}, \frac{9}{2}, \frac{23}{2}, \frac{25}{2},
\frac{39}{2}, \frac{41}{2}\hbar$ in band $\nu7/2^{+}[633]$, which
are listed in Table~\ref{Table:2}. The orbital $|4\rangle$, i.e.,
$7/2^{+}[633]$ in Table~\ref{Table:1}, is the dominant component at
the bandhead with $ \sum_{K} | C_{4}^{IK}|^2 \sim75\%$. As the
nucleus rotates faster, due to the large Coriolis matrix elements of
$i_{13/2}$ sub-shell, at $I=\frac{41}{2}$, the contribution of the
orbital $|4\rangle$ is decreased to $\sim 21\%$.

\subsection{Band $\nu5/2^{-}[512]$}

\begin{center}
\begin{figure*}[h!]
\centering
\includegraphics[width=16cm]{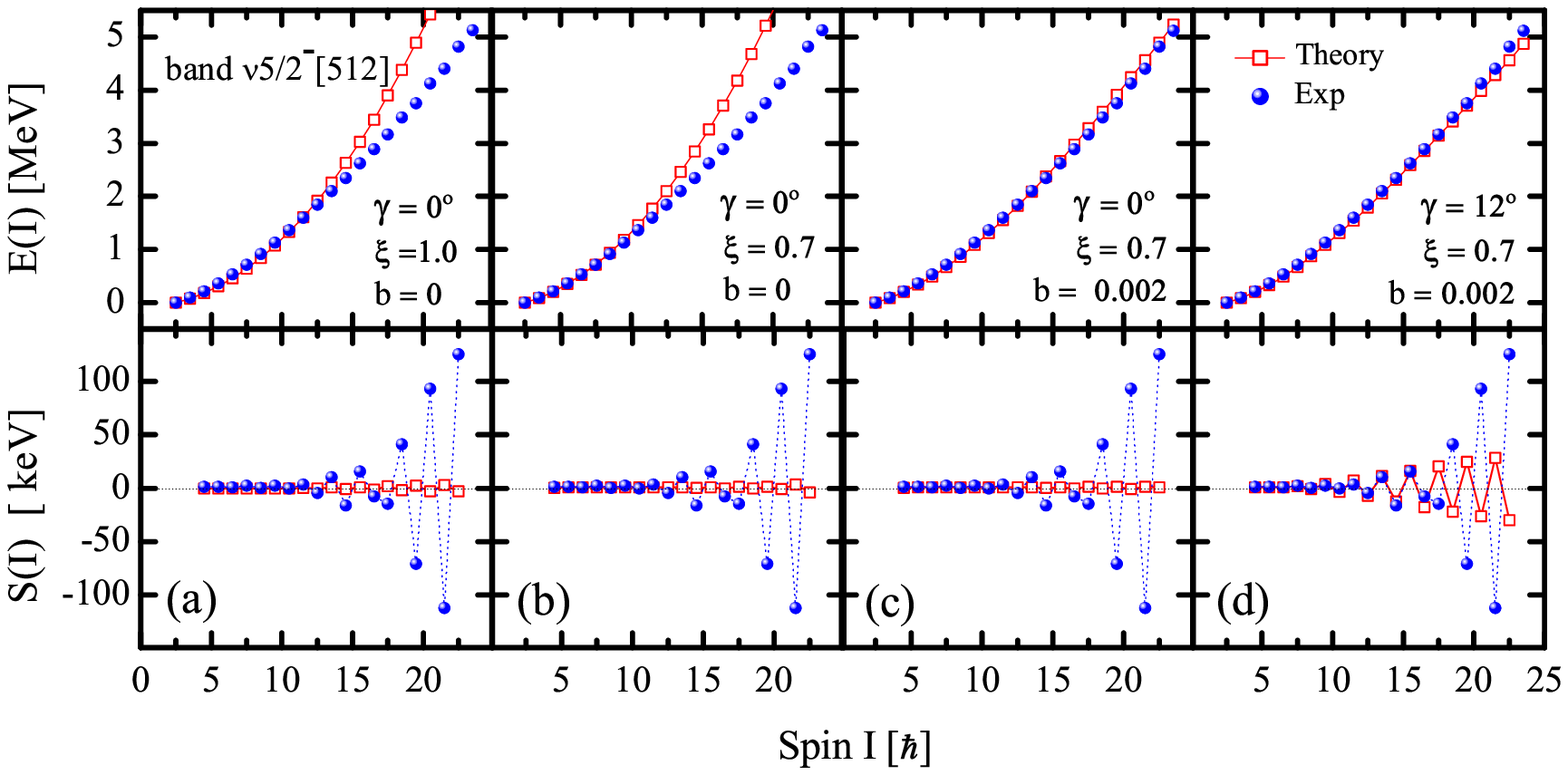}
\caption{(color online) The energy spectra $E(I)$ and signature
splitting $S(I)$ for the band $\nu5/2^{-}[512]$ in $^{173}$W based
on the PRM (red squares) in comparison with the data (blue
points)~\cite{ZhangYH06}. In the calculations, $\varepsilon_2=0.24$,
and ${\cal J}_0 = 35$ MeV$^{-1}\hbar^2$. The pairing parameter
$\Delta_{n} = 0.81$ MeV. The Fermi energy $\lambda_n$, triaxial
deformation parameter $\gamma$, Coriolis attenuation factor $\xi$
and  the parameter $b$ in the variable moment of inertia in the
panels are respectively, (a) $\lambda_n = 51.41$ MeV,
$\gamma=0^{\circ}$, $\xi=1.0$ and $ b=0 $; (b) $\lambda_n = 51.41$
MeV, $\gamma=0^{\circ}$, $\xi=0.7$ and $ b=0 $; (c) $\lambda_n =
51.41$ MeV, $\gamma=0^{\circ}$, $\xi=0.7$ and $ b=0.002 $; (d)
$\lambda_n = 51.51$ MeV, $\gamma=12^{\circ}$, $\xi=0.7$ and $
b=0.002$.
 }\label{fig:fig4}
\end{figure*}
\end{center}

The energy spectra $E(I)$ and signature splitting $S(I)$ for the
band $\nu5/2^{-}[512]$ calculated by the PRM are given in
Fig.~\ref{fig:fig4} and compared with data~\cite{ZhangYH06}. In the
calculations, $\varepsilon_2=0.24$, and ${\cal J}_0 = 35$
MeV$^{-1}\hbar^2$. The pairing parameter $\Delta_{n} = 0.81$ MeV.
The Fermi energy $\lambda_n$, triaxial deformation parameter
$\gamma$, Coriolis attenuation factor $\xi$ and the parameter $b$ in
the variable moment of inertia in the panels are respectively, (a)
$\lambda_n = 51.41$ MeV, $\gamma=0^{\circ}$, $\xi=1.0$ and $ b=0 $;
(b) $\lambda_n = 51.41$ MeV, $\gamma=0^{\circ}$, $\xi=0.7$ and $ b=0
$; (c) $\lambda_n = 51.41$ MeV, $\gamma=0^{\circ}$, $\xi=0.7$ and
$b=0.002 $; (d) $\lambda_n = 51.51$ MeV, $\gamma=12^{\circ}$,
$\xi=0.7$ and $ b=0.002$.

Using a constant moment of inertia, the calculated energy can only
agree with data in the lower spin region as shown in
Fig.~\ref{fig:fig4}a. As shown in Fig.~\ref{fig:fig4}b, compared
with the band $\nu7/2^{+}[633]$, the Coriolis attenuation has less
influence for the band $\nu5/2^{-}[512]$ due to the low $j$
components ($\frac{7}{2}$ and $\frac{9}{2}$). To reproduce the
observed energy spectra $E(I)$, a variable moment of inertia is
necessary, as shown in Fig.~\ref{fig:fig4}c.

Similar to the band $\nu7/2^{+}[633]$, without the $\gamma$ degree
of freedom, the calculated amplitudes of signature splitting $S(I)$
are small compared with the observed ones. As shown in
Fig.~\ref{fig:fig4}d, after introducing the triaxiality ($\gamma =
12^{\circ}$), signature splitting $S(I)$ for $I < \frac{35}{2}\hbar$
is reproduced perfectly. Without invoking the triaxiality,
systematic calculations by simply adjusting all the other
parameters, including $\varepsilon_2$ and $\Delta$, have been done
and it is found that the spectra $E(I)$ and the signature splitting
$S(I)$ can not be reproduced simultaneously for band
$\nu5/2^{-}[512]$, which shows the necessary of triaxial deformation
for this band.

Although signature splitting $S(I)$ for $I < \frac{35}{2}\hbar$ can
be well reproduced by introducing triaxiality, the calculated
signature splitting $S(I)$ for $I \geq \frac{35}{2}\hbar$ is
inconsistent with the data. In particular, there is a signature
inversion at spin $\frac{35}{2}\hbar$, i.e., $S(I)$ is smaller for
$I=2n+1/2$ ($\alpha=+1/2$) for $I<\frac{35}{2}\hbar$ and for
$I=2n-1/2$ ($\alpha=-1/2$) for $I\geq\frac{35}{2}\hbar$, which
cannot be reproduced by the present calculation.

The main components of the wave functions for spins $I =
\frac{5}{2}, \frac{7}{2}, \frac{23}{2}, \frac{25}{2}, \frac{39}{2},
\frac{41}{2}\hbar$ in band $\nu 5/2^-[512]$ are listed in
Table~\ref{Table:3} to examine the evolution of the wave function.
The orbital $|3\rangle$, i.e., $5/2^-[512]$ in Table~\ref{Table:1},
is the dominant component at the bandhead  with $ \sum_{K} |
C_{3}^{IK}|^2 \sim 96 \%$. As the nucleus rotates faster, at
$I=\frac{41}{2}$, the contribution of the orbital $|3\rangle$ is
decreased to $\sim 40\%$. As given in Table~\ref{Table:1}, both
orbitals $2f_{7/2}$ and $1h_{9/2}$ contribute significantly to the
orbital $|3\rangle$, i.e., $5/2^-[512]$. As the orbital $2f_{7/2}$
contributes $\sim 64\%$ and the orbital $1h_{9/2}$  $\sim 25\%$ to
the orbital $|3\rangle$, we attempt to understand
Fig.~\ref{fig:fig4}d in terms of the orbitals $2f_{7/2}$ and
$1h_{9/2}$ in the following.

\begin{center}
\begin{figure*}[h!]
\centering
\includegraphics[width=8cm]{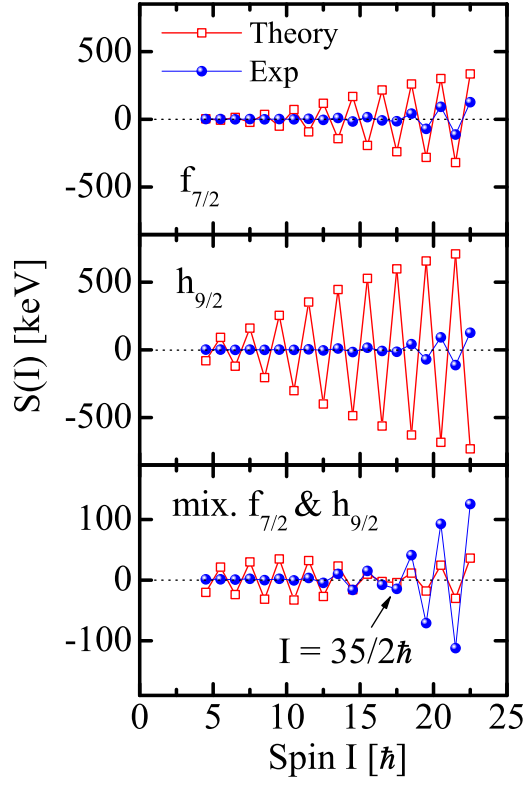}
\caption{(color online) Signature splitting $S(I)$ in PRM
calculations (red squares) with the valence neutron in $2f_{7/2}$
single-$j$ shell (upper panel), $1h_{9/2}$ single-$j$ shell (middle
panel) and their mixture (lower panel) in comparison with the data
(blue points) for the band $\nu5/2^{-}[512]$ in
$^{173}$W~\cite{ZhangYH06}. In the calculations,
$\varepsilon_2=0.24$, $\gamma=12^{\circ}$, ${\cal J}_0=35$
MeV$^{-1}\hbar^2,~b=0.002$, $\xi=1.0$, $\Delta_n$=0.81 MeV, and
$\lambda_{n}$ is taken to be the energy of the level
$\Omega=\frac{5}{2}$ in the corresponding single-$j$ shell.
}\label{fig:fig5}
\end{figure*}
\end{center}

For the band with one quasiparticle in a high-$j$ orbital, the
favored signature can be obtained by
$\alpha_f=(-1)^{j-1/2}1/2$~\cite{Stephens75}, i.e., $\alpha_f=-1/2$
for orbital $2f_{7/2}$ and $\alpha_f=+1/2$ for orbital $1h_{9/2}$.
This is consistent with the calculated results in
Fig.~\ref{fig:fig5}, in which the signature splitting $S(I)$ in PRM
calculations with the valence neutron in $2f_{7/2}$ single-$j$ shell
(upper panel), $1h_{9/2}$ single-$j$ shell (middle panel) and their
mixture (lower panel) are given in comparison with the data for the
band $\nu5/2^{-}[512]$  in $^{173}$W~\cite{ZhangYH06}. In the
calculations, $\varepsilon_2=0.24$, $\gamma=12^{\circ}$, ${\cal
J}_0=35$ MeV$^{-1}\hbar^2,~b=0.002$, $\xi=1.0$, $\Delta_n$=0.81 MeV,
and $\lambda_{n}$ is taken to be the energy of the level
$\Omega=\frac{5}{2}$ in the corresponding single-$j$ shell.

For $I < \frac{35}{2}\hbar$, the observed phase of $S(I)$ in band
$\nu5/2^{-}[512]$ is consistent with that for $1h_{9/2}$
configuration, but different from that for $2f_{7/2}$. For $I \geq
\frac{35}{2}\hbar$, the observed phase of $S(I)$ are no longer
consistent with that for $1h_{9/2}$, but consistent with that for
$2f_{7/2}$. As the amplitudes of $S(I)$ are concerned, the
amplitudes of $S(I)$ for $1h_{9/2}$ are larger than those for
$2f_{7/2}$ and both of them are much larger than the observed ones.
Therefore, although the orbital $2f_{7/2}$ is the predominant
component in the observed band $\nu5/2^{-}[512]$, the strong
competition from $1h_{9/2}$ results in signature splitting $S(I)$
for the $I<\frac{35}{2}\hbar$ as shown in Fig.~\ref{fig:fig4}d.

As for signature inversion at $I = \frac{35}{2}\hbar$, one
straightforward explanation may be attributed to the superiority of
the orbital $2f_{7/2}$ over $1h_{9/2}$ after $I >
\frac{35}{2}\hbar$. By mixing the energy spectra according to the
mixing ratios of $2f_{7/2}$ and $1h_{9/2}$ in orbital
$\nu5/2^{-}[512]$ in Table.~\ref{Table:1}, the $S(I)$ is extracted
and compared with the observed one in band $\nu5/2^{-}[512]$ in the
lower panel in Fig.~\ref{fig:fig5}. It is interesting to note that
the observed phase of $S(I)$ and signature inversion in band
$\nu5/2^{-}[512]$ can be perfectly reproduced although the
deficiencies still exist for the amplitude. We also note that there
is an upbending for this band around $I=\frac{35}{2}\hbar$ based on
the analysis of spin alignment and the moment of inertia. However, a
self-consistent description of the upbending and signature inversion
is beyond the scope of the present one quasiparticle PRM.

\section{Conclusion}

A particle rotor model with a quasi-neutron coupled with a triaxial
rotor is applied to study signature splitting of the bands built on
the intruder orbital $\nu7/2^{+} [633]$ and non-intruder orbital
$\nu5/2^{-}[512]$ in $^{173}$W.  With triaxiality and Coriolis
attenuation, the correct spin of the lowest state,  energy spectra
$E(I)$ and signature splitting $S(I)$ for the band $\nu7/2^{+}[633]$
are well reproduced.

The phase and amplitude of signature splitting in the band of the
non-intruder orbital $\nu5/2^{-}[512]$ is attributed to the strong
competition of the contribution from its main components $2f_{7/2}$
and $1h_{9/2}$. Similar competitions should take important roles in
bands with other non-intruder orbitals, and the conclusion is also
helpful to interpret the signature splitting in the neighbor odd-odd
nuclei, such as the band $\pi h_{9/2}\otimes\nu 5/2^- [512]$ in
$^{174}$Re~\cite{ZhangYHCPL07}. Although triaxiality ($\gamma =
12^{\circ}$) can reproduce signature splitting $S(I)$ for $I <
\frac{35}{2}\hbar$ and energy spectra $E(I)$ perfectly, signature
inversion for $I = \frac{35}{2}\hbar$ in band $\nu5/2^{-}[512]$
still cannot be understood by the triaxiality, Coriolis attenuation,
and variability of the moment of inertia. As an upbending happens, a
proper description of signature inversion self-consistently is
beyond the scope of the present one quasiparticle PRM.

\begin{acknowledgments}
Helpful discussions with G.C. Hillhouse, Y.H. Zhang and X.H. Zhou
are gratefully acknowledged. This work is partly supported by Major
State Basic Research Developing Program 2007CB815000, the National
Natural Science Foundation of China under Grant Nos. 10505002,
10435010, 10605001, 10775004 and 10221003.
\end{acknowledgments}

\newpage

\begin{table}[h!]
\begin{center}
\tabcolsep=10pt \caption{Positive and negative single-particle
levels $|\nu\rangle$ adopted for the band $\nu7/2^+[633]$ at
$\gamma=15^\circ$, and the band $\nu5/2^-[512]$ at $\gamma=12^\circ$
for the present triaxial PRM calculations. The approximate Nilsson
quantum numbers, single particle energies, and main components
expanded in the basis $|Nlj\Omega\rangle$ are shown. The neutron
Fermi energy $\lambda_n$ is $51.46$ MeV for $\gamma=15^\circ$, and
$51.51$ MeV for $\gamma=12^\circ$. }
\begin{tabular}{ccccl}
 \hline\hline
 $\gamma$    & $|\nu\rangle$ & $\Omega^\pi[Nn_{z}\Lambda]$ & $\varepsilon_\nu$ (MeV)&    Main components in terms of $|Nlj\Omega\rangle$ \\
 \hline
             &  $|1\rangle$ & ~$1/2^+[660]$            &  49.32      & $0.784|6i_{13/2}~\,\frac{1}{2}\rangle ~+ 0.423|6i_{13/2}\textendash\frac{3}{2}\rangle ~ + 0.353 |6g_{9/2}~~\,\frac{1}{2}\rangle$ \\
             &  $|2\rangle$ & ~$3/2^+[651]$            &  50.16      & $0.798|6i_{13/2}\textendash\frac{3}{2}\rangle ~- 0.360|6i_{13/2}~\,\frac{1}{2}\rangle ~+ 0.337 |6g_{9/2}~\textendash\frac{3}{2}\rangle$\\
             &  $|3\rangle$ & ~$5/2^+[642]$            &  50.90      & $0.903|6i_{13/2}~\,\frac{5}{2}\rangle ~-0.292|6g_{9/2}~~\,\frac{5}{2}\rangle ~+ 0.182 |6g_{9/2}~~\,\frac{1}{2}\rangle$\\
 $15^{\circ}$&  $|4\rangle$ & ~$7/2^+[633]$            &  51.95      & $0.951|6i_{13/2}\textendash\frac{7}{2}\rangle ~ +0.214|6 g_{9/2}~\textendash\frac{7}{2}\rangle ~- 0.159|6g_{9/2}~\textendash\frac{3}{2}\rangle $\\
             &  $|5\rangle$ & ~$9/2^+[624]$            &  53.26      & $0.973|6i_{13/2}~\,\frac{9}{2}\rangle ~+0.157|6g_{9/2}~~\,\frac{5}{2}\rangle ~-0.124|6g_{9/2}~~\,\frac{9}{2}\rangle$\\
             &  $|6\rangle$ & ~$1/2^+[651]$            &  53.85      & $0.566|6g_{9/2}~~\,\frac{1}{2}\rangle ~-0.531|6d_{5/2}~~\,\frac{1}{2}\rangle ~-0.238|6s_{1/2}~~\,\frac{1}{2}\rangle $\\
             &  $|7\rangle$ & $11/2^+[615]$            &  54.76      & $0.985|6i_{13/2}\textendash\frac{11}{2}\rangle-0.156|6g_{9/2}~\textendash\frac{7}{2}\rangle~+0.060|6i_{11/2}\textendash\frac{11}{2}\rangle$\\
 \hline
             &  $|1\rangle$ & ~$3/2^-[521]$            &  49.78      & $0.670|5f_{7/2}\textendash\frac{3}{2}\rangle~ - 0.592|5h_{9/2}\textendash\frac{3}{2}\rangle$ ~+ 0.221$|5p_{3/2}\textendash\frac{3}{2}\rangle$\\
             &  $|2\rangle$ & ~$5/2^-[523]$            &  50.26      & $0.814|5h_{9/2}~\,\frac{5}{2}\rangle~ - 0.394|5f_{7/2}~\,\frac{5}{2}\rangle~+ 0.227|5f_{5/2}~\,\frac{5}{2}\rangle$\\
             &  $|3\rangle$ & ~$5/2^-[512]$            &  51.51      & $0.798|5f_{7/2}~\,\frac{5}{2}\rangle~+ 0.495|5h_{9/2}~\, \frac{5}{2}\rangle ~- 0.213|5p_{3/2}~\,\frac{1}{2}\rangle$\\
 $12^{\circ}$&  $|4\rangle$ & ~$1/2^-[521]$            &  51.63      & $0.465|5p_{1/2}~\,\frac{1}{2}\rangle~ +0.457|5f_{7/2}~\,\frac{1}{2}\rangle ~-0.442|5h_{9/2}~\, \frac{1}{2}\rangle $\\
             &  $|5\rangle$ & ~$7/2^-[514]$            &  52.08      & $0.922|5h_{9/2}\textendash\frac{7}{2}\rangle~+ 0.274|5f_{7/2}\textendash\frac{7}{2}\rangle~- 0.195|5f_{5/2} \textendash\frac{3}{2}\rangle$\\
             &  $|6\rangle$ & ~$7/2^-[503]$            &  53.50      & $0.841|5f_{7/2}\textendash\frac{7}{2}\rangle~- 0.340|5h_{9/2}\textendash\frac{7}{2}\rangle ~- 0.264|5p_{3/2}\textendash\frac{3}{2}\rangle$\\
             &  $|7\rangle$ & ~$1/2^-[510]$            &  53.56      & $0.500|5f_{5/2}\textendash\frac{3}{2}\rangle~- 0.468|5p_{3/2}~\,\frac{1}{2}\rangle ~ - 0.406|5f_{5/2}~\,\frac{1}{2}\rangle$\\
 \hline
 \hline
\end{tabular}\label{Table:1}
\end{center}
\end{table}

\begin{table}[h!]
\begin{center}
\tabcolsep=12pt \caption{The main components expanded in the strong
coupling basis $|IMK\nu\rangle$ in Eq.(\ref{eq:base}) (denoted as $
|K \nu \rangle$ for short) for selected states in the band
$\nu7/2^{+}[633]$. The parameters in the calculations are the same
as in Fig.~\ref{fig:fig1}c. }
\begin{tabular}{clll}
\hline\hline
 $I^\pi$             & Main components in terms of $|K \nu\rangle$\\
 \hline
 $\frac{7}{2}^+$     & $0.868|\textendash\frac{7}{2}\,4\rangle  + 0.485|~\,\frac{5}{2}\,3\rangle - 0.094|\textendash\frac{3}{2}\,2\rangle - 0.019|\textendash\frac{3}{2}\,3\rangle$\\
 $\frac{9}{2}^+$     & $0.806|\textendash\frac{7}{2}\,4\rangle  - 0.554|~\,\frac{5}{2}\,3\rangle + 0.143|\textendash\frac{3}{2}\,2\rangle - 0.113|~\,\frac{9}{2}\,5\rangle$\\
 $\cdots$            & $\cdots$ \\
 $\frac{23}{2}^+$    & $0.652|\textendash\frac{7}{2}\,4\rangle + 0.582|~\,\frac{5}{2}\,3\rangle + 0.232|\textendash\frac{3}{2}\,2\rangle + 0.224|~\,\frac{9}{2}\,5\rangle$ \\
 $\frac{25}{2}^+$    & $0.558|~\,\frac{5}{2}\,3\rangle  - 0.500 |\textendash\frac{7}{2}\,4\rangle - 0.371|\textendash\frac{3}{2}\,2\rangle - 0.283|~\,\frac{1}{2}\,3\rangle$ \\
 $\cdots$            & $\cdots$ \\
 $\frac{39}{2}^+$    & $0.524|\textendash\frac{7}{2}\,4\rangle  - 0.474|~\,\frac{5}{2}\,3\rangle - 0.263|~\,\frac{9}{2}\,3\rangle  - 0.256|\textendash\frac{3}{2}\,4\rangle$\\
 $\frac{41}{2}^+$    & $0.465|~\,\frac{5}{2}\,3\rangle - 0.418|\textendash\frac{3}{2}\,2\rangle  - 0.340|\textendash\frac{7}{2}\,4\rangle - 0.334|~\,\frac{1}{2}\,3\rangle$\\
 $\cdots$            & $\cdots$ \\
 \hline\hline
\end{tabular}
\label{Table:2}
\end{center}
\end{table}

\begin{table}[h!]
\begin{center}
\tabcolsep=12pt \caption{Same as Table.~\ref{Table:2}, but for
selected states in the band $\nu5/2^{-}[512]$ and the parameters in
the calculations are the same as Fig.~\ref{fig:fig4}d.}
\begin{tabular}{cl}
 \hline\hline
  $I^\pi$            & Main components in terms of $| K \nu\rangle$\\
  \hline
  $\frac{5}{2}^-$    & $0.982|\frac{5}{2}\,3\rangle - 0.111|~\,\frac{5}{2}\,4\rangle ~+0.088|\textendash\frac{3}{2}\,1\rangle - 0.078|\textendash\frac{3}{2}\,4\rangle $\\
  $\frac{7}{2}^-$    & $0.926|\frac{5}{2}\,3\rangle + 0.313|\textendash\frac{7}{2}\,5\rangle ~-0.126|\textendash\frac{3}{2}\,1\rangle +0.074|~\,\frac{5}{2}\,4\rangle $\\
  $\cdots$           & $\cdots$ \\
  $\frac{23}{2}^-$   & $0.669|\frac{5}{2}\,3\rangle + 0.487|\textendash\frac{7}{2}\,5\rangle ~-0.228|~\,\frac{9}{2}\,3\rangle -0.207|~\,\frac{1}{2}\,3\rangle  $ \\
  $\frac{25}{2}^-$   & $0.583|\frac{5}{2}\,3\rangle - 0.464|\textendash\frac{7}{2}\,5\rangle ~-0.231|~\,\frac{1}{2}\,3\rangle -0.219|~\,\frac{9}{2}\,3\rangle  $ \\
  $\cdots$           & $\cdots$ \\
  $\frac{39}{2}^-$   & $0.532|\frac{5}{2}\,3\rangle + 0.408|\textendash\frac{7}{2}\,5\rangle ~-0.326|~\,\frac{9}{2}\,3\rangle - 0.301|~\,\frac{1}{2}\,3\rangle $ \\
  $\frac{41}{2}^-$   & $0.449|\frac{5}{2}\,3\rangle - 0.374|\textendash\frac{7}{2}\,5\rangle ~-0.277|~\,\frac{1}{2}\,3\rangle - 0.276|~\,\frac{9}{2}\,3\rangle $ \\
  $\cdots$           & $\cdots$ \\
 \hline\hline
\end{tabular}
\label{Table:3}
\end{center}
\end{table}

\end{document}